\documentclass[12pt,dvips]{elsart}
\usepackage[dvips]{graphicx}
\usepackage{amssymb}
\begin{document}
\newcommand{\arcdeg}{\hbox{$^\circ$}}
\newcommand{\arcmin}{\hbox{$^\prime$}}
\newcommand{\arcsec}{\hbox{$^{\prime\prime}$}}
\newcommand{\simless}{\,\kern.1em\lower.5ex\hbox{$\buildrel <
\over\sim$}\kern.1em\,}
\newcommand{\simgreat}{\,\kern.1em\lower.5ex\hbox{$\buildrel >
\over\sim$}\kern.1em\,}
\runauthor{Kawachi}
\begin{frontmatter}
\title{The optical reflector system for the CANGAROO-II 
 imaging atmospheric Cherenkov telescope}
\author[ICRR]{A.~Kawachi\thanksref{e-mail}}
\author[TITech]{Y.~Hayami},
\author[Tokai]{J.~Jimbo},
\author[TITech]{S.~Kamei},
\author[ICRR]{T.~Kifune},
\author[TITech]{H.~Kubo},
\author[TITech]{J.~Kushida},
\author[ICRR]{S.~LeBohec\thanksref{stephan_next}},
\author[Mitsubishi]{K.~Miyawaki},
\author[ICRR]{M.~Mori},
\author[Tokai]{K.~Nishijima},
\author[Adelaide]{J.R.~Patterson},
\author[TITech]{R.~Suzuki},
\author[TITech]{T.~Tanimori},
\author[Ibaraki]{S.~Yanagita},
\author[ICRR]{T.~Yoshikoshi\thanksref{yoshikoshi_next}},
\author[STE]{A.~Yuki}
\thanks[e-mail]{E-mail: kawachi@icrr.u-tokyo.ac.jp}
\thanks[stephan_next]{present address: 
 Department of Physics and Astronomy, Iowa State University, 
 Ames, IA 50011-3160, U.S.A.}
\thanks[yoshikoshi_next]
{present address: Department of Physics, Osaka City University, 
 Sumiyoshi-ku, Osaka 558-8585, Japan}
\address[ICRR]{Institute for Cosmic Ray Research, University of Tokyo, 
 Tanashi, Tokyo 188-8502, Japan}
\address[TITech]{Department of Physics, Tokyo Institute of Technology, 
        Meguro, Tokyo 152-8551, Japan}
\address[Tokai]{Department of Physics, Tokai University, 
 Hiratsuka, Kanagawa 259-1292, Japan}
\address[Mitsubishi]{Communication Systems Center, 
 Mitsubishi Electric Corporation, Amagasaki, Hyogo 661-8661, Japan}
\address[Adelaide]{Department of Physics and Mathematical Physics, University of Adelaide, South Australia 5005, Australia}
\address[Ibaraki]{Faculty of Science, Ibaraki University, 
   Mito, Ibaraki 310-8521, Japan}
\address[STE]{STE Laboratory, Nagoya University,
   Nagoya, Aichi 464-8601, Japan}
\begin{abstract}
A new imaging atmospheric Cherenkov telescope
with a light-weight reflector has been constructed.
Light, robust, and durable mirror facets of 
containing CFRP (Carbon Fiber Reinforced Plastic) laminates were developed 
for the telescope.  
The reflector has a parabolic shape ($f$/1.1) with a 
30~m$^2$ surface area which consists of sixty spherical mirror facets.
The image size of each mirror facet is 0$\arcdeg$.08 (FWHM) on average.
The attitude of each  facet can be adjusted by stepping motors.
After the first in situ adjustment, a point image of 
 about 0$\arcdeg$.14 (FWHM) over 3 degree field-of-view was obtained.
The effect of gravitational load on the optical system 
was confirmed to be negligible at the focal plane.
The telescope has been in operation 
with an energy threshold for gamma-rays of $\simless$300~GeV since May 1999.
\end{abstract}
\end{frontmatter}

\section{Introduction}
\label{sect:intro}

 It is in the last decade that 
 major discoveries and progress in the field of TeV gamma-ray
 astronomy have been achieved by using ground-based imaging 
 atmospheric Cherenkov telescopes\,\cite{akawachi:vhe_gamma_review1,akawachi:vhe_gamma_review2,akawachi:vhe_gamma_review3}.
 Currently, the lowest achieved energy threshold among the imaging 
 Cherenkov telescopes is about 250\,GeV.
 The experimental technique is considered to be still in its 
 development phase, and we need improvements for 
 the next generation of imaging atmospheric Cherenkov telescopes.
 At present, the number of reported TeV sources is about 10,
 while more than 250 sources are reported in the 100\,MeV to 
 30\,GeV band in the third EGRET catalogue\,\cite{akawachi:EGRET_catalogue}.
 So, a wealth of results is expected in the unexplored energy 
 region near 100\,GeV, and world-wide efforts are being 
 made to lower the energy threshold of detectable gamma-rays 
 down to 100\,GeV or even below so that the energy coverage 
 overlaps with that of satellite detectors; 
 for example, 
 the STACEE\,\cite{akawachi:STACEE} and 
 CELESTE\,\cite{akawachi:CELESTE} experiments which have started 
 operations using solar heliostats as large area photon collectors, 
 the MAGIC project\,\cite{akawachi:MAGIC} 
 for a high sensitivity telescope of a 17-m diameter mirror,
 the VERITAS project\,\cite{akawachi:VERITAS_proposal}
 and the HESS project\,\cite{akawachi:HESS_proposal} for arrays of 
 multiple imaging telescopes of 10-m aperture size.
 The CANGAROO project 
 has studied TeV gamma-ray sources since 1992, 
 at Woomera, South Australia (136$\arcdeg$\,E, 31$\arcdeg$ S, 220\,m a.s.l.) 
 with a 3.8-m imaging atmospheric Cherenkov 
 telescope\,\cite{akawachi:CANGAROO-I-1,akawachi:CANGAROO-I-2,akawachi:CANGAROO-III}.
 In the second phase  of the project (CANGAROO-II), the 3.8-m telescope 
 has been replaced by a 
 7-m telescope located near the 3.8-m telescope site which 
 started operation in May 1999\,\cite{akawachi:CANGAROO-II-1,akawachi:CANGAROO-II-2}.
 The third phase of the CANGAROO project (CANGAROO-III) has commenced, 
 and an array of four 10-m telescopes 
 will start observation in 2004\,\cite{akawachi:CANGAROO-III}.

 A typical imaging atmospheric Cherenkov telescope has a reflector to 
 collect Cherenkov light from extensive air showers which develop in the upper 
 atmosphere, and to focus the showers' images onto a multi-pixel camera. 
 The images have an extended shape, governed by 
 the Cherenkov angle and by multiple scattering of the charged 
 shower components.
 The average Cherenkov photon density is almost proportional to 
 the primary energy of the incident gamma-ray.
 Thus, a larger reflector 
 can directly lower the observational gamma-ray energy threshold.
 Gamma-ray showers from a point-like source are discriminated from the 
 overwhelming background of cosmic ray showers 
 using differences in their images caused by their differing 
 shower developments\,\cite{akawachi:image_analysis}.
 The analyses of the image shapes of the showers 
 with an accuracy of 0$\arcdeg$.1--0$\arcdeg$.2 
 has led to the success of the imaging Cherenkov technique in 
 TeV gamma-ray detection
 (\,\cite{akawachi:vhe_gamma_review2,akawachi:vhe_gamma_review3} and references therein).
 Within the constraints of 
 engineering, fabrication, and assembly costs, 
 telescopes are usually limited to diameters of 10\,m, and 
 the large light collection area is obtained by tessellated 
 multiple mirror facets of diameter less than 1\,m.
 As the reflector size increases, 
 the gravitational load on the optical system 
 may cause pointing deviations 
 or deformation of images at the focal plane 
 during observations.
 The optimum design has to attempt to minimize the effect of 
 gravity within the other constraints on the design and construction.

 For the 7-m reflector, 
 we have developed a new type of mirrors based on 
 CFRP (Carbon Fiber Reinforced Plastic) laminate.
 Our trial for mirrors of composite materials like CFRP encountered 
 some difficulties in their realization;  
 production of a composite mirror facet of different materials 
 with good accuracies  requires  
 delicate control over many parameters, e.g. slight changes of 
 autoclave temperature as a function of time.
 Nevertheless, the merit of a CFRP laminate mirror is 
 that it is light, robust, and durable.
 After developing the CFRP-based mirror,
 we have been able to construct a large but light-weight and 
 easy to handle optical system almost free from the effect of gravity, 
 at reasonable cost and with manageable construction labor.
 We also employed computer-controlled 
 stepping motors for on-site alignment of each mirror facet 
 which enables us to obtain the best focus of the 7-m tessellated reflector.
 The telescope is also a prototype for, and will be the first telescope 
 of, the CANGAROO-III project.

 In this paper, we describe the design,  manufacturing, 
 alignment procedure, 
 and the performance of the CANGAROO-II optical reflector system,
 expanding upon the preliminary descriptions given 
 elsewhere\,\cite{akawachi:kawachi_opt1,akawachi:kawachi_opt2}.
 All of these techniques provide us with useful  knowledge 
to prepare for final design of the CANGAROO-III 
 telescope as well as of even larger Cherenkov telescopes for 
 the future.

\section{Design of the 7-m Telescope}
\label{sect:design}

\begin{figure}[ht] 
\begin{center}
\includegraphics[width=14.cm,keepaspectratio]{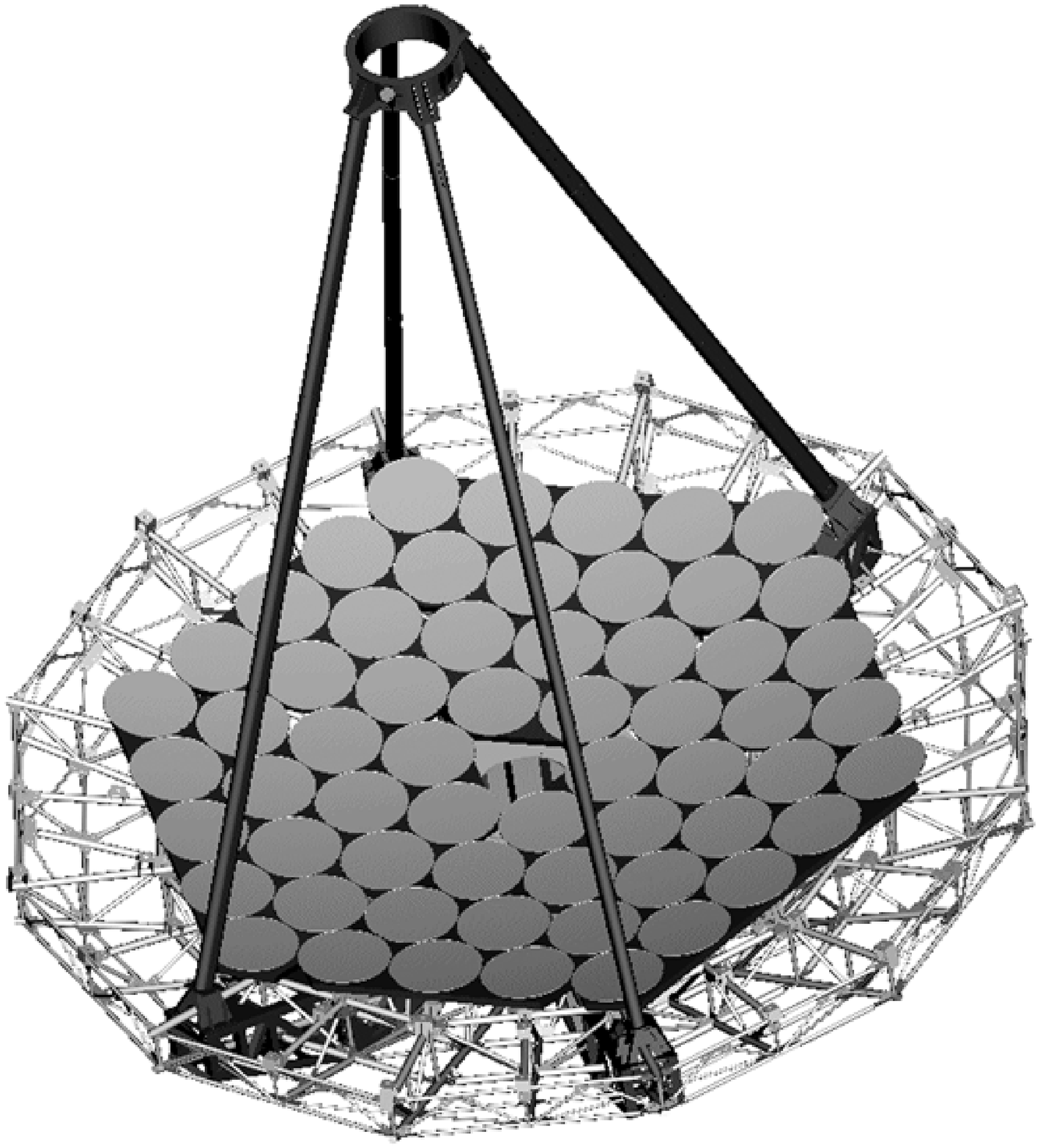}
\caption{
 A schematic drawing of 
 the reflector  and the camera support of the CANGAROO-II telescope.
 The paraboloid reflector is 7.2\,m  in diameter 
 (an effective area of 30\,m$^2$)  with an $f$-number of 1.1.
 The dish consists of sixty mirror facets, nine honeycomb panels 
 for the mirror mounts, and the reflector trusses.
 The motor-driven system for alignment is installed under each facet.}
\label{fig:akawachi:whole}
\end{center}
\end{figure}

 Figure\,\ref{fig:akawachi:whole} shows a view of the reflector 
 and the camera support of the telescope. 
 The reflector is an $f$/1.1 tessellated paraboloid with a diameter of 
 7\,m, and with an effective light collecting area of 30\,m$^2$.
 Each of the sixty mirror facets is spherical in shape, and has 
 an 0.8\,m diameter and a radius of curvature of 16.4\,m (on average).
 The prime focal plane of the reflector is equipped with 
 a camera of 512 fast photo-multiplier tubes (PMTs) of 
 13\,mm diameter with UV-glass (Hamamatsu, R4124UV).
 The PMTs are arranged with a spacing of 16\,mm (0$\arcdeg$.115) 
 to cover a field-of-view of about 3$\arcdeg$, 
 and are supplemented by 16\,mm\,$\times$\,16\,mm  light-collecting cones 
 to reduce the dead space between photosensitive area of the 
 PMTs.

\subsection{Mechanical Structure}

 The reflector trusses 
 were based on the technique of commercially available radio communication 
 antennas\footnote{Mitsubishi Electric Corporation, 
 Communication Systems Center}, 
 and the nine honeycomb panels are mounted on the trusses.
 Six to nine mirror facets were installed on each honeycomb panel. 
 The rough alignment of the mirror facets was performed 
 at the factory for each panel, which were then shipped 
 with the facets installed.
 The camera support has a simple mechanical structure 
 where the four camera stays are connected directly to 
 the center ring to hold the camera.
 The present support frame allows us to extend the reflector to 
 10-m diameter by the additional of 54 more mirror facets
 early in the year 2000.

 The total weight of the moving part of the 7-m telescope is 
 6.3\,ton, similar to that of of the 6\,ton CAT telescope\,\cite{akawachi:cat}.
 However, the 7-m telescope has almost twice the effective reflector area
 of the CAT telescope.
 The light-weight but robust structure 
 was designed to be able to be operated 
 in winds of average velocity 30\,km/hr 
 and operational for gusts up to 100\,km/hr.
 The load of the camera and its cables 
 is about 100\,kg at the 8-m focal length.
 Gravitational deflections of the camera support
 are a potential cause of focal point shifting, however  
 the deviation in the pointing accuracy 
 at all the elevation angles was measured to be 
 less than 1$\arcmin$ (the nominal value).
 These  measurements will be presented in more detail 
 in Section\,\ref{sect:result_grav}.

\subsection{Optical Parameters}
\label{sect:optical_parameters}

 Gamma-ray showers can be preferentially selected 
 over accidental events due to the night sky background by 
 the use of a narrow timing gate 
 of 10\,ns or so\,\cite{akawachi:cangaroo_design}.
 For good timing information,  we chose a 
 paraboloid design which provides isochronous 
 collection of photons. 
 The alternative design used 
 in a number of other imaging atmospheric Cherenkov 
 telescopes (for example, see\,\cite{akawachi:whipple_mirror_study}) 
 is the Davies-Cotton type of reflector 
 where identical spherical mirror facets 
 are mounted on a spherical structure with a radius of curvature 
 that is exactly half that of the facets\,\cite{akawachi:davies-cotton}.
 The maximum variation in photon arrival times 
 from different portions of a Davies-Cotton type ($f$/0.7)
 10-m reflector is 6\,ns\,\cite{akawachi:whipple_mirror_study}, while 
 the variation calculated with our tessellated paraboloid design 
 is less than 0.2\,ns, even for an extended diameter of 10\,m.

\begin{figure}[ht] 
\begin{center}
\vspace*{50pt}
\includegraphics[width=12.cm]{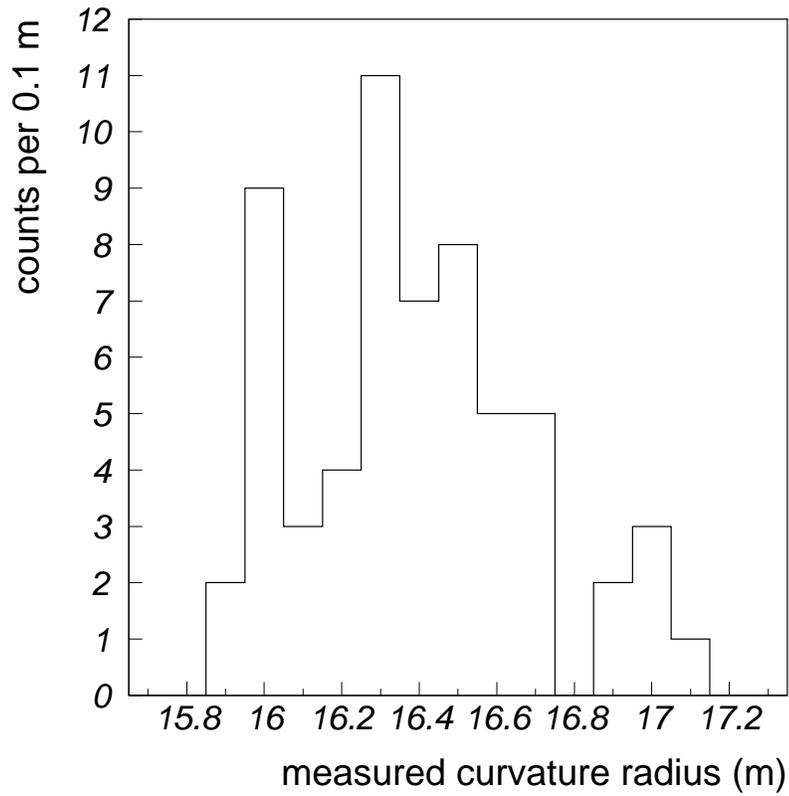}
\caption{The distribution of measured radii of curvature of 
 the sixty mirror facets. The measurement error was 0.1\,m.}
\label{fig:akawachi:curvature_radii}
\end{center}
\end{figure}

 Another important point in the optical designing was 
 the off-axis focusing ability; it is desirable to minimize the 
 smearing of the point spread function and resulting off-axis 
 decrease in the light concentration.
 Generally, a spherical reflector 
 has better off-axis properties than a paraboloid, 
 and the Davies-Cotton design has an advantage over the paraboloid design 
 for Cherenkov imaging telescopes which require a wide field-of-view
 (usually more than 3 degrees).
 When appropriate radii of curvature of the facets are chosen, 
 a parabolic tessellated reflector 
 can achieve acceptable off-axis performance, especially in 
 the light concentration.
 As the result of a ray-tracing calculation, 
 for a 7-m reflector of sixty spherical mirror facets, 
 a 16.4\,m radius of curvature gives the best performance when 
 all facets have identical curvature radii\,\cite{akawachi:cangaroo_design}.
 An arrangement of the mirror facets of slightly differing 
 radii of curvature 
 improves the performance further. 
 There is a variance in the  radii of curvature 
 of the CFRP composites in the manufacturing process 
 (Figure\,\ref{fig:akawachi:curvature_radii}) 
 and we took advantage of this by arranging the 
 facets based on the radii of curvature referring to 
 the results  of the  ray-tracing program.

 Assuming the point spread function of a facet ($\sigma = {\rm 0}\arcdeg$.1),
 the light concentration within one camera pixel (0$\arcdeg$.115 square) 
 was calculated by a ray-tracing program with our design parameters, 
 and with the Davies-Cotton design of the same $f$-number, 
 of the same size and number of facets.
 The centers of off-axis images are shifted on the focal plane 
 by the combined effect of the blurs of the facets and 
 their off-axis aberrations.
 The ray-tracing calculations estimated the peak-point shifts of 
 about 5\,\% (our design) and of 1\,\% (the Davies-Cotton design)
 at the edge of our field-of-view (1$\arcdeg$.3 off the axis).
 On-axis and at the edge of the field-of-view,
 the portions of photons which fall within a camera pixel 
 at the image center are estimated to be 
 59\,\% and 48\,\% 
 for the paraboloid, 
 59\,\% and 51\,\% 
 for the Davies-Cotton type. 
 The absolute performances of the two designs differ only 3\,\% at most.

 In summary, our design offers good on-axis performance with adequate 
 off-axis focusing and achieves our goal of a photon arrival time spread 
 of 1\,ns.
 The off-axis performance of the 7-m reflector has been directly measured
 and will be reported  in 
 Section\,\ref{sect:result_off}.

\section{Spherical Mirror Facets}
\label{sect:facets}

\begin{figure}[ht] 
\begin{center}
\vspace*{150pt}
\includegraphics[width=14.cm,keepaspectratio]{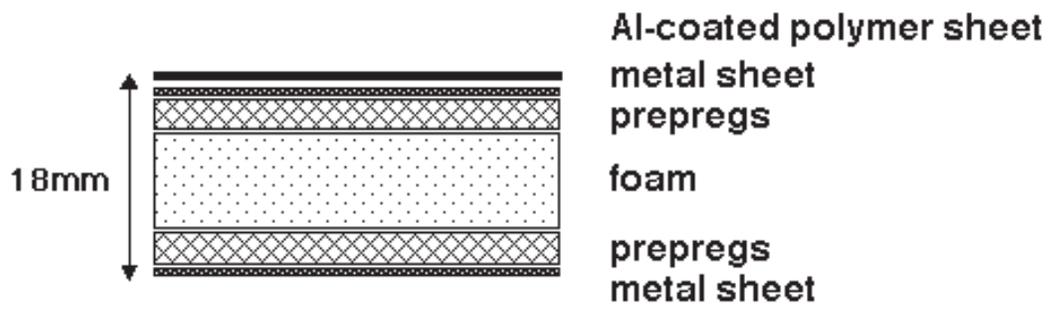}
\caption{ A schematic illustration of the cross section of 
 a spherical mirror facet.
 The ``prepreg'' is the intermediate composite form made by 
 carbon fibers impregnated with resin. 
}
\label{fig:akawachi:crosssecion}
\end{center}
\end{figure}

 Each spherical mirror facet is  0.8\,m in diameter, 18\,mm thick, 
 and weighs only 5.5\,kg. 
 The average density of the facet  is about one fifth of 
 the ordinary glass (2.4--2.6\,g$\cdot$cm$^{-3}$), and for example,  
 the CAT telescope employs
 a spherical mirror facet made of borosilicate glass 
 which is 0.5\,m in diameter and 10\,mm thick\,\cite{akawachi:cat}.
 A schematic cross section of a facet is shown in
 Figure\,\ref{fig:akawachi:crosssecion}. 
 Prepregs (sheets of carbon fiber impregnated with resin)
 were laid on a metal mold, 
 sandwiching a thick core of low density, high shear-strength foam 
 to avoid twisting deformations.
 A sloped edge was premachined on the core to achieve a better figure after 
 being shaped.
 The radius of curvature of the mold was 16.4\,m.
 Deflections of the facet  
 by gravity were estimated to be as small as 
 a few\,$\mu$m at the edge of a facet.
 A polymer sheet coated with laminated aluminum
 was applied on the top of these layers as the reflecting material.
 The facets were then sealed and cured to 120$\arcdeg$C 
 in an autoclave pressure vessel.
 No mechanical polishing of the surface was carried out.

 Fiber patterns of the prepreg materials may make the mirror surface 
 less smooth,  and may  cause a considerable amount of random scattering 
 of incident lights.
 The possible scattering loss was investigated 
 by a CCD measurement of the light focused by the CFRP mirror 
 and by a glass mirror of good accuracy as a reference.
 Although the test was subject to non-negligible errors, 
 it was confirmed that at least 91\,\% of the light is 
 focused by the CFRP mirror (without a reflectivity 
 calibration of two materials).

 We tested possible deterioration of the facets 
 by repeating 200 cycles of a change in temperature between 
 0$^\circ$ and 50$^\circ$C in 2 hours.
 Changes of the humidity up to 90\% were also included in the test.
 Within measurement errors of 0.1\,m, no change in the curvature 
 was found after this test.

\begin{figure}[ht] 
\begin{center}
\vspace*{-10pt}
\includegraphics[width=14.cm,keepaspectratio]{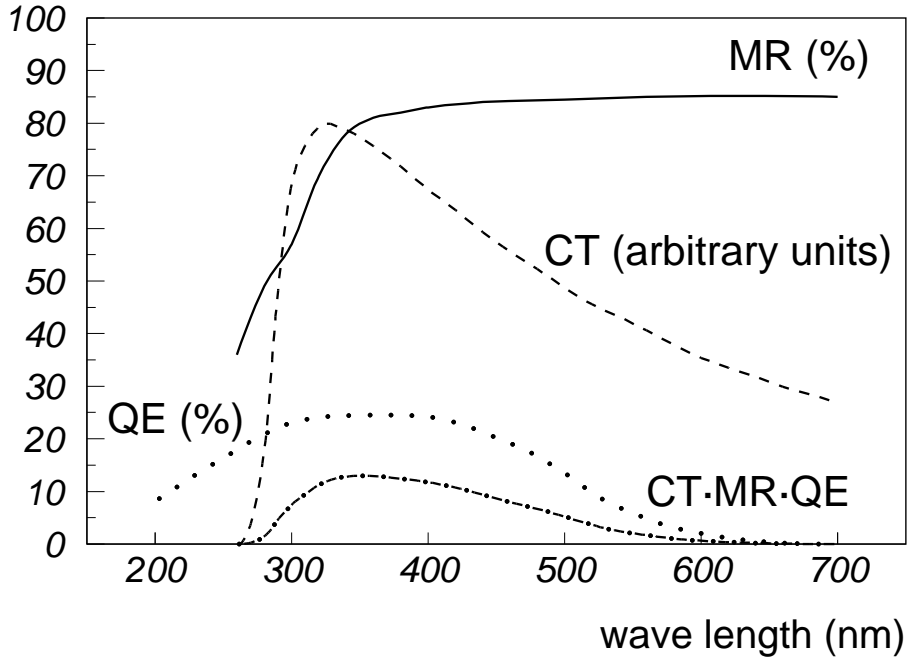}
\caption{
 The mirror reflectivity after the surface coating (MR) is 
 shown as a function of wavelength (solid line).
 The quantum efficiency (QE) of R4124UV cathode (taken from the 
 Hamamatsu catalogue) is also plotted (dotted line),
 together with 
 the Cherenkov emission spectrum after atmospheric transmission (CT) 
 in arbitrary units (dashed line).
 This spectrum multiplied by MR and QE (CT$\cdot$MR$\cdot$QE) is 
 shown as the dot-dashed line.
 The transmission spectrum was taken 
 from the figure in \protect\,\cite{akawachi:transmission}.} 
\label{fig:akawachi:ref}
\end{center}
\end{figure}

 The telescope is not sheltered, so the facets are exposed to 
 tough environmental conditions.
 For protection against dust, rain, and sunshine, 
 the facets were coated with fluoride.
 Figure\,\ref{fig:akawachi:ref} shows the measured reflectivity 
 as a function of wavelength (solid line).
 The reflectivity is over 80\,\% at 340\,nm--800\,nm and 
 falls off rather slowly to 40\,\% at 250\,nm, 
 corresponding both to the 
 atmospheric transmission cut-off of Cherenkov light at 300\,nm 
 (the Cherenkov light spectrum after 
 transmission\,\cite{akawachi:transmission} is 
 shown by the dashed line) 
 and to the response of the PMT photo-cathode with a UV-transparent 
 window (dotted line).
 The reflectivity of the reflector on-site is monitored by a hand-held 
 reflectometer at the wavelength of 480\,nm\,\cite{akawachi:reflectometer}.
 We found dust on the surface reduces the reflectivity to 
 about 75\,\% after several months, 
 however, we confirmed with a sample that over a year 
 the reflectivity repeatedly recovers easily by washing with water.
 The surface is free from dew
 until the relative humidity exceeds 83\,\% 
 when the wind is not strong.

\begin{figure}[ht] 
\begin{center}
\vspace*{50pt}
\includegraphics[width=12.cm,keepaspectratio]{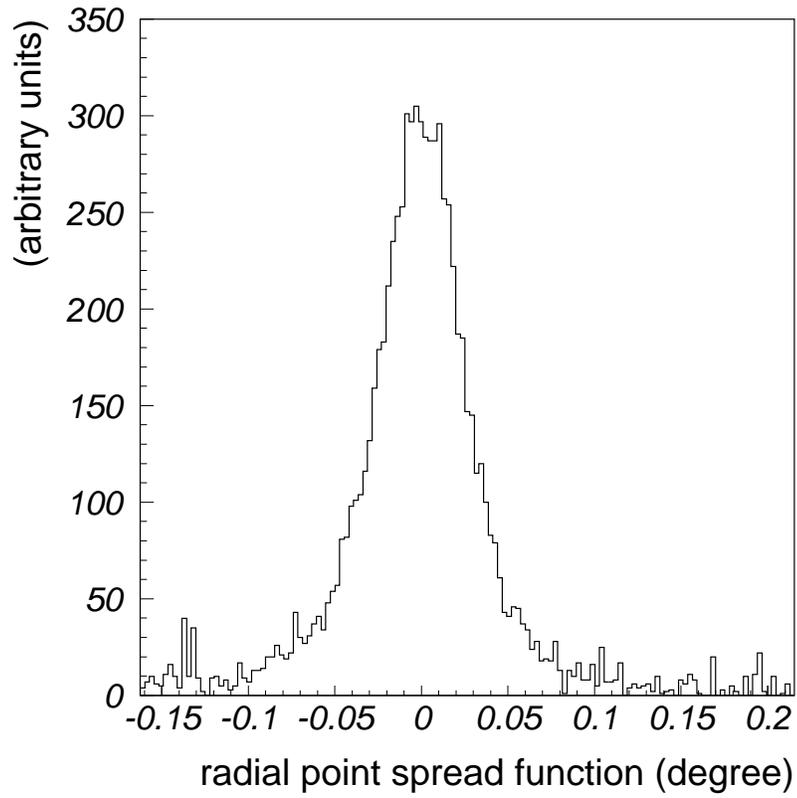}
\caption{A slice of the CCD image of one mirror facet 
 with a distant light source.
 The radial point spread function of this mirror is 0$\arcdeg$.06 (FWHM).
 }
\label{fig:akawachi:good_1d}
\end{center}
\end{figure}

 Using an approximate-point light source, 
 every spherical mirror facet was examined at the factory, and 
 its focal length and point spread function were measured.
The absolute values of the curvature radii were calibrated 
by the sampling measurements  
of three-dimensional configuration of some facets.
The image size of each mirror facet 
was also measured on-site with a light source 5.8\,km away.
The radii of curvature of the sixty mirrors are distributed between 
15.9--17.1\,m (Figure\,\ref{fig:akawachi:curvature_radii}), 
 with an average of 16.4\,m and a root-mean-square of 0.3\,m.
 The permissible variation 
 was estimated by a ray-tracing calculation.
 The facets were arranged according to their radii of curvature 
 from the inner to the outer sections of the reflector, 
 with the shorter-curvature radii ones innermost.
 A CCD image of one mirror (with a distant light source) 
 is shown in Figure\,\ref{fig:akawachi:good_1d}.
 A typical point spread function of the sixty facets is 0$\arcdeg$.08 (FWHM), 
 and 50\,\% and 80\,\% of the photons are 
 concentrated within 0$\arcdeg$.1 and 0$\arcdeg$.15 circles, 
 respectively.

\section{Mirror Alignment by the Motor-Driven System}
\label{sect:adjustment}

\begin{figure}[ht] 
\begin{center}
\vspace*{150pt}
\includegraphics[width=15.cm,keepaspectratio]{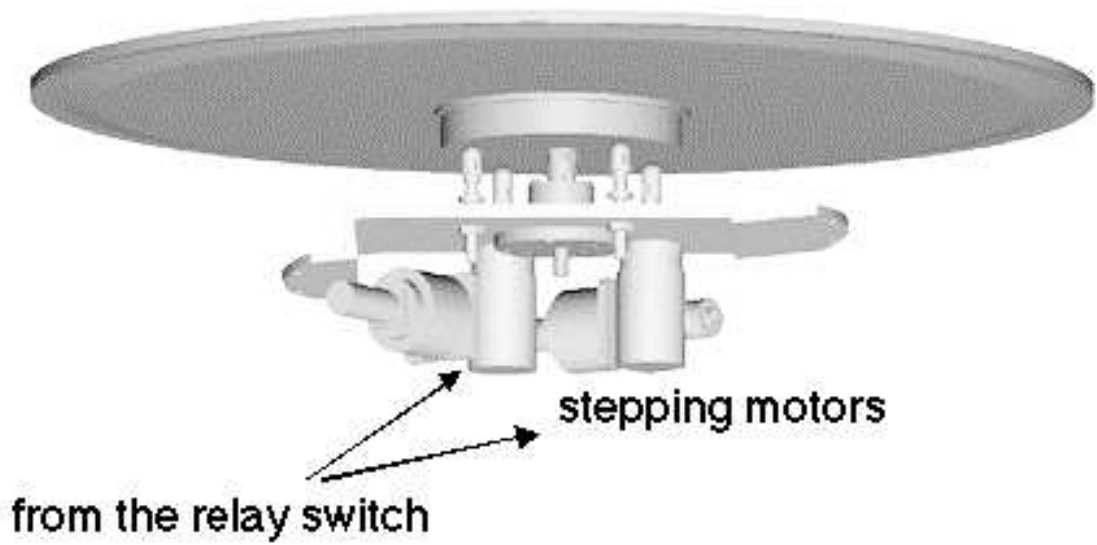}
\caption{
 The motor-driven system for alignment of one mirror facet is 
 schematically illustrated. 
 Four shafts support a stainless boss at the back surface of each facet,  
 where two shafts are driven by two stepping motors, and the other two shafts 
 with springs fix the attitude.
 Facets are adjusted one by one using two motor drivers with relay switches
 controlled by a computer.
}
\label{fig:akawachi:motor}
\end{center}
\end{figure}

 Two watertight stepping motors are installed at the back 
 surface of each mirror facet, 
 and the attitude can be remotely adjusted 
 in two perpendicular directions (Figure\,\ref{fig:akawachi:motor}).
 A stainless boss at the back surface of each facet 
 is connected to the honeycomb panel 
 by a universal joint, and four shafts support the boss.
 Two shafts are driven by two stepping motors, and the other two shafts 
 with springs firmly fix the attitude.
 The minimum step size corresponds to about 1$\times{\rm 10}^{-4}$ degree 
 at the focal plane, and the facets are adjustable up to $\pm$3~degrees.
 An accuracy of 1$\times{\rm 10}^{-3}$ 
 degree is retained when motors are switched off.
 Facets are adjusted one by one using two motor drivers with relay switches
 controlled by a computer.

 At the factory,  the alignment of the facets was roughly adjusted 
 to within 0$\arcdeg$.3 relative to each supporting panel 
 using a laser beam.
 With these alignment works,  we were able to check our 
 stepping-motor system and save on-site labor.

 The alignment on-site used the 5.8\,km distant light source at night. 
 All mirror facets but one were covered with plastic lids,  
 and the image of the uncovered facet was monitored 
 on a screen at the focal plane by a CCD camera installed at the 
 center of the dish. 
 The attitudes of the facets were adjusted 
 with the stepping motors using feedback information from the 
 CCD images so that the image center lay at the focal point of the 
 reflector.

 The use of approximate-parallel light during the adjustment work
 may have caused a systematic 
 shifting of the focal length of the reflector. 
 To examine this, after all the sixty facets were adjusted,  
 the effective focal length was measured with a star image.
 By moving the focal plane along the optical axis 
 we surveyed for the optimum point which was found to coincide with the 
 design focal length of 8.00\,m within an error of 0.01\,m.
 As a result of  the initial adjustment work, the facet axes deviated by 
 0$\arcdeg$.01--0$\arcdeg$.07.

\section{Performance of the Tessellated Reflector}
\label{sect:result}

 The optical properties of the reflector as a whole
 were measured using images of 
 several stars tracked by the telescope.
 Images on the focal plane screen were taken by a CCD camera at the reflector 
 center.  

\subsection{On-axis Properties}
\label{sect:result_on}

\begin{figure}[ht] 
\begin{center}
\includegraphics[width=14.cm,keepaspectratio]{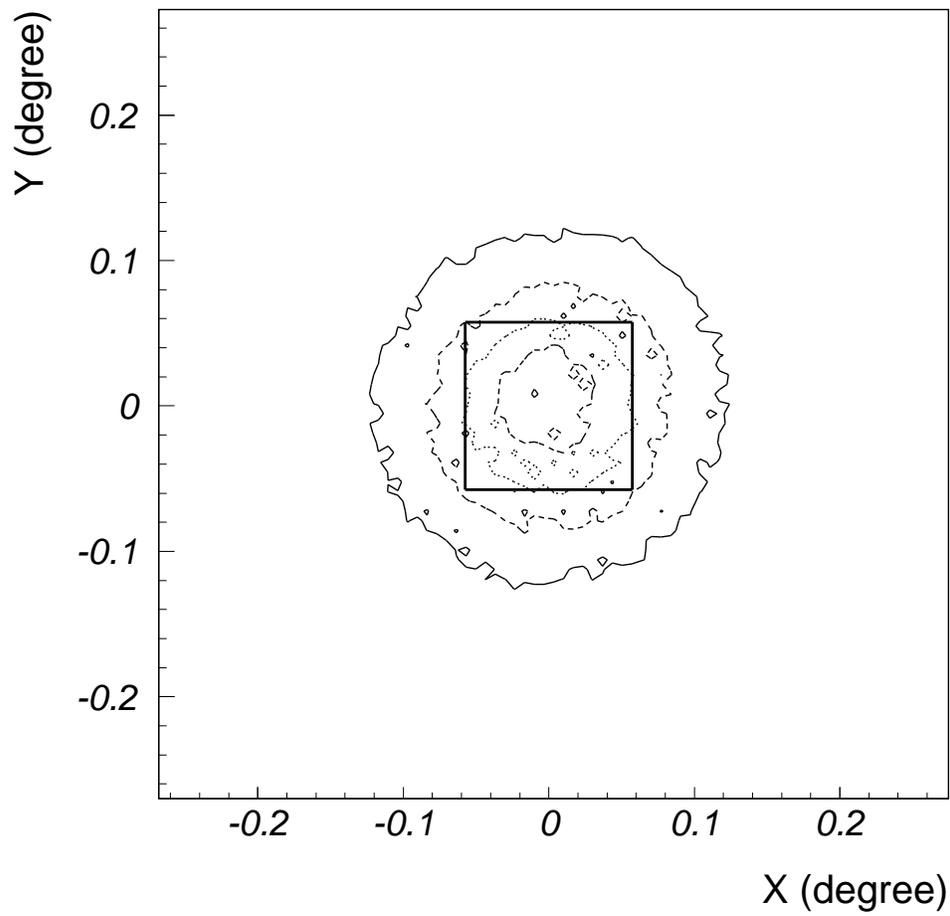}
\caption{A CCD image of Sirius on the optic axis. 
 The contours are in steps of 20\,\% of the peak intensity.
 A square is superimposed to 
 show the scale of  a camera pixel (0$\arcdeg$.115 square).
}
\label{fig:akawachi:on_axis}
\end{center}
\end{figure}

 Figure\,\ref{fig:akawachi:on_axis} shows an on-axis image of 
 Sirius (visual magnitude of $-$1.5).
 One camera pixel (0$\arcdeg$.115 square) is superimposed for scaling.
 An image size of 0$\arcdeg$.14\,$\pm$\,0$\arcdeg$.01 (FWHM) was deduced; 
 30\,$\pm$\,4\,\% of the photons are concentrated in a single camera pixel, 
 and 50\,\% of the photons are 
 concentrated within a circle of 0$\arcdeg$.16\,$\pm$\,0$\arcdeg$.02 in 
 diameter.
 As described in Section\,\ref{sect:intro},  
 the characteristic difference in size between the images of gamma-rays and 
 protons is 0$\arcdeg$.1--0$\arcdeg$.2.
 Our optical quality meets the requirement, although 
 we are planning to improve the alignment.

 The relativistic charged particles in extensive air showers 
 emit the Cherenkov radiation.  Only  high energy muons in the showers 
 are likely to reach the ground without decaying and 
 individual muons will radiate a characteristic Cherenkov ring  
 (the Cherenkov threshold energy is about 4.4\,GeV).
 These rings are detected in the focal plane as arcs or rings, 
 depending on the impact parameters.
 A preliminary analysis of observed images shows thin muon rings 
 with an average width of about 0$\arcdeg$.11\,\cite{akawachi:okumura}.
 For $\sim$10\,GeV muons, 
 the contribution of reflector aberrations to 
 the broadening of the ring images can be estimated 
 to be approximately 1.5 times the contribution of 
 the multiple scattering in the 
 atmosphere\,\cite{akawachi:whipple_muon_study}.

\subsection{Off-axis Properties}
\label{sect:result_off}

\begin{figure}[ht] 
\begin{center}
\includegraphics[width=14.cm,keepaspectratio]{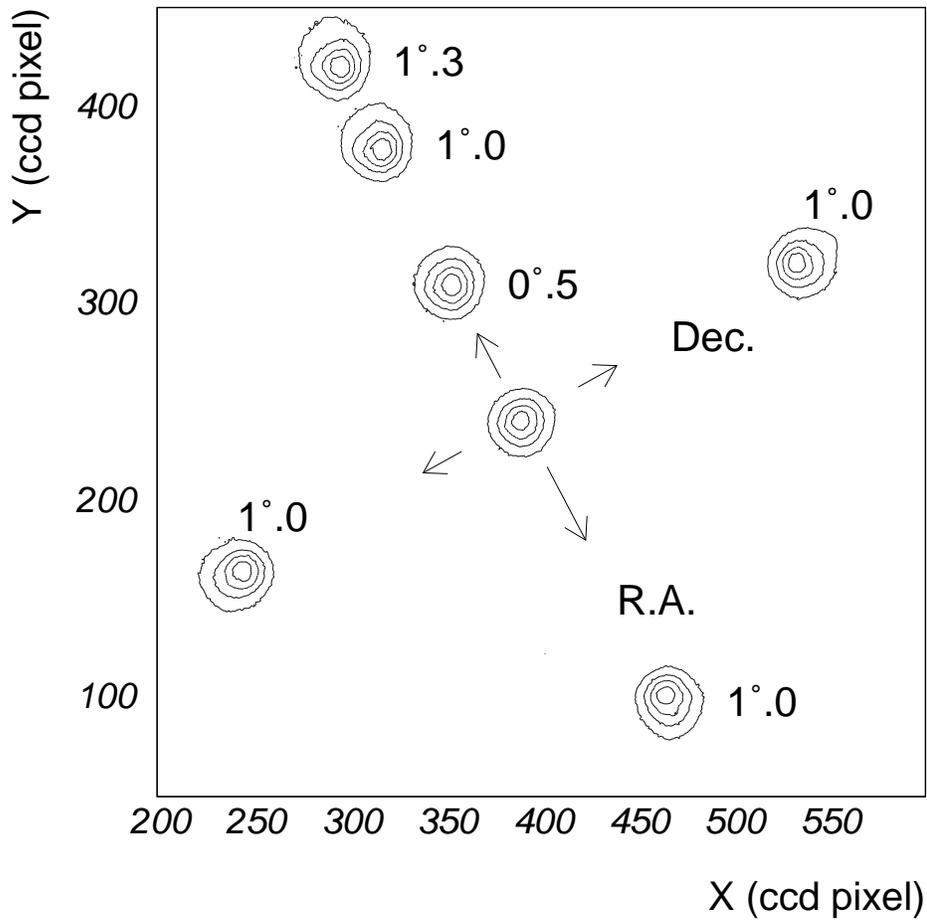}
\caption{A synthesized figure of some of the CCD images 
  obtained by displacing the telescope pointing from a star.
  The axes are in unit of CCD pixels, 
  corresponding to 6.7$\times$10$^{-3}$ degree. 
  The contours are in steps of 20\,\% of the peak intensity of 
  of each image.}
\label{fig:akawachi:off_axis1}
\end{center}
\end{figure}

\begin{figure}[ht] 
\begin{center}
\includegraphics[width=14.cm,keepaspectratio]{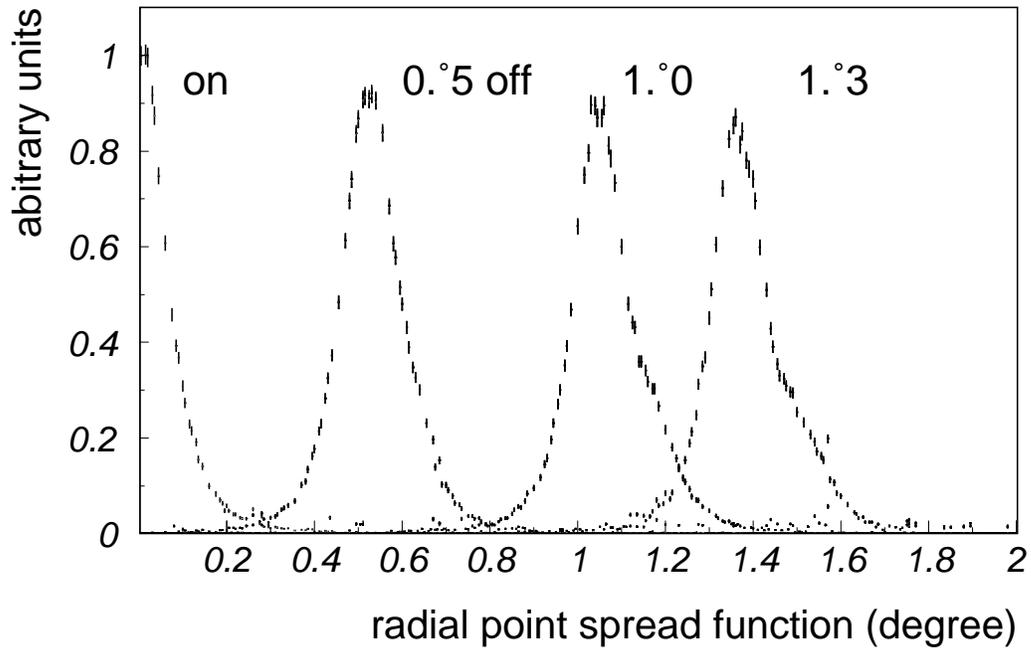}
\vspace*{10pt}
\caption{
 Radial point spread functions at the  different 
 pointing coordinates;  ($\alpha$,\,$\delta$),
 ($\alpha-$0$\arcdeg$.5,\,$\delta$)
 ($\alpha-$1$\arcdeg$.0,\,$\delta$)
 and  ($\alpha-$1$\arcdeg$.3,\,$\delta$), respectively, where 
 $\delta =$ ($-$16$\arcdeg$\,42$\arcmin$\,58$\arcsec$ (J2000)).
 Vertical scales are normalized by the peak height of the 
 on-axis point spread function.}
\label{fig:akawachi:off_axis2}
\end{center}
\end{figure}

 We compared off-axis images of Sirius 
 displacing the telescope pointing in both 
 right ascension and declination.
 Some of the CCD images are synthesized and 
 shown in Figure\,\ref{fig:akawachi:off_axis1}.
 Figure\,\ref{fig:akawachi:off_axis2}  shows 
 radial point spread functions of the star 
 displaced in right ascension  by 
 0$\arcdeg$.5, 1$\arcdeg$.0 and 1$\arcdeg$.3, respectively.
 Vertical scales are normalized by the peak height of the 
 on-axis function.
 The light concentration  within a camera pixel (0$\arcdeg$.115 square) 
 are  28\,\%, 26\,\%, and 25\,\% (the errors are 3\,\% for all 
 three), compared to the on-axis value of  30\,$\pm$\,4\,\%  mentioned earlier.
 In comparison, the ray-tracing program predicted a 
 relative decrease to 81\,\% of the on-axis concentration
 at 1$\arcdeg$.3 off axis.
 The peak centers were shifted as a result of the convolution of 
 blurs of the facets and 
 the aberrations (Section\,\ref{sect:optical_parameters}).
 The position of the peak center is, for example, 
 1$\arcdeg$.43 after the scale calibration, 
 by averaging 
 the displacements of 1$\arcdeg$.3 to the four directions.

\section{Effect of Gravitational Deformations}
\label{sect:result_grav}

 At the different attitudes during 
 observations, pointing deviations or deformation of 
 images may appear at the focal plane due to 
 gravitaional effects; such as deflection of the facets, 
 of the camera stays, and of the reflector trusses.
 Furthermore, 
 the facet alignment was performed while the telescope 
 was pointed horizontally, thus 
 the fixed setting may also have deflection to be calibrated.
 By tracking various stars, the effects were directly measured 
 as the CCD images on the focal plane screen.

\begin{figure}[ht] 
\begin{center}
\includegraphics[width=14cm,keepaspectratio]{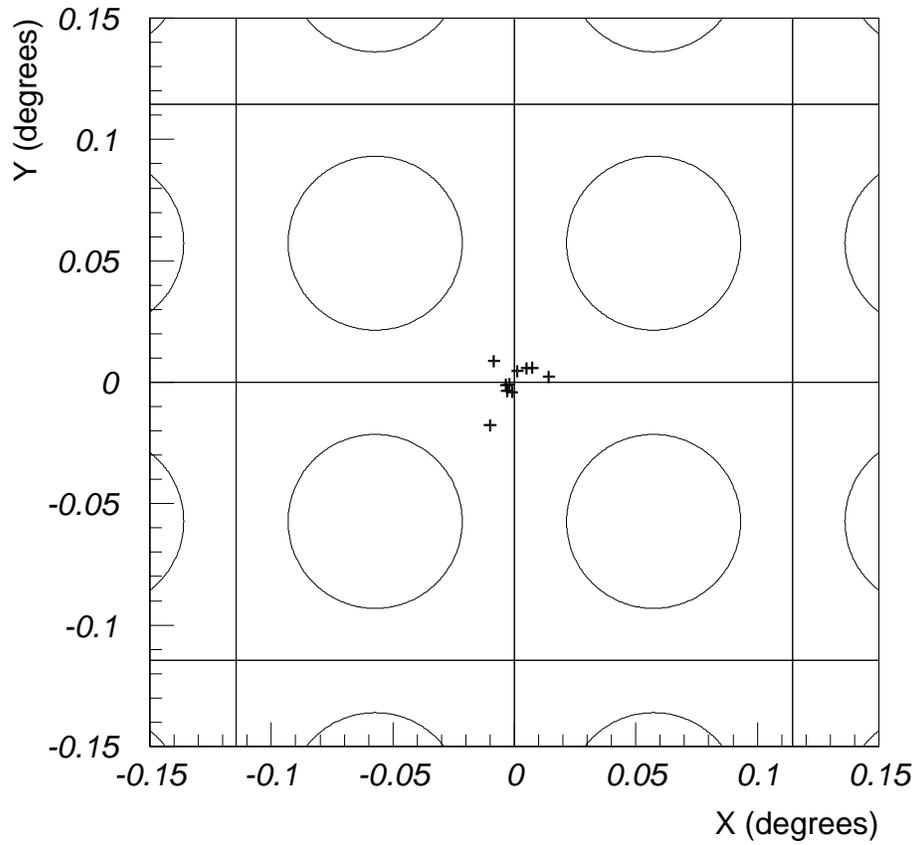}
\caption{ 
 A scatter plot of the center positions of the star images 
 taken at elevations between 12$\arcdeg$ and 85$\arcdeg$ and 
 over all azimuthal angles.
 The origin of the figure is the focal point of the telescope.
 An array of camera pixels (0$\arcdeg$.115 square) and of 
 photocathode area (0$\arcdeg$.09 $\phi$) of PMTs are 
 superimposed.
}
\label{fig:akawachi:pointing_accuracy1}
\end{center}
\end{figure}

 At elevations between 12$\arcdeg$ and 85$\arcdeg$ and over 
 all the azimuthal angles,  
 the pointing had a root-mean-square of 27$\arcsec$ 
 (Figure\,\ref{fig:akawachi:pointing_accuracy1}), 
 with no dependence 
 on elevation or azimuthal angles.

\begin{figure}[ht] 
\begin{center}
\includegraphics[width=14.cm,keepaspectratio]{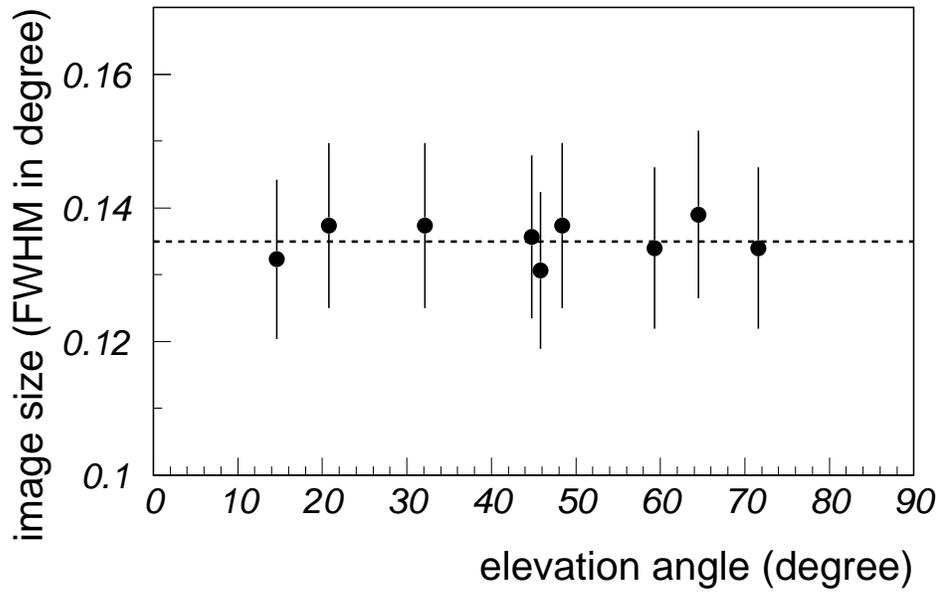}
\caption{Measured size (FWHM) of star images 
 on the focal plane taken  
 at elevation angles between 15$\arcdeg$ and 70$\arcdeg$.
 The average value (0$\arcdeg$.135) of the stars 
 is indicated as a dashed line.
}
\label{fig:akawachi:el_dep}
\end{center}
\end{figure}

 Figure\,\ref{fig:akawachi:el_dep} shows the image sizes of 
 nine stars as a function of the elevation angle.
 For elevation angles between 15--70 degrees,  
 the images show no dependence on elevation in size within a 
 measurement error of 0$\arcdeg$.01.
 The deformation of the image shape is confirmed to be also negligible 
 The fitted results 
 of the two-dimensional point spread function of the images 
 coincide well with each other, and  
 the average eccentricity of the nine images 
 is 0.99\,$\pm$\,0.01\, (statistical error only).

\section{Summary}
 The new CANGAROO-II 7\,m telescope has been constructed to 
 observe the southern sky for very high-energy gamma-rays 
 sources in the sub-TeV region.
 Observation has started with the telescope, 
 which has a light collection area of 30\,m$^2$ and 
 and an energy threshold of less than 300\,GeV.
 The $f$/1.1 paraboloid reflector offers nearly isochronous timing, 
 and the curvature radii of facets were chosen to improve the off-axis 
 focusing.
 The light-weight optical system of CFRP facets 
 allowed considerable savings in construction labor,  
 while the gravitational torques on the system are 
 negligible at the focal plane. 
 The facets are easy to handle and easy to maintain even in the open air 
 environment.
 The motor-driven alignment of the tessellated reflector on-site
 has been shown to work well with gains in efficiency and safety.
 After our initial alignment was completed, 
 the optical performance of the reflector was examined using star images.
 The point spread function is 0$\arcdeg$.14 (FWHM), comparable in size 
 to a camera pixel of the camera (0$\arcdeg$.115 square).
 About 30\,\% of the photons of an on-axis star image fall within 
 a single camera pixel. The light concentration is reduced 
 by only 18\,\% at the edge of the 3$\arcdeg$ field-of-view, 
 confirming that 
 off-axis aberrations are not severe for this telescope design.

\section{Acknowledgments}
 The authors thank P.G.~Edwards for careful reading of the manuscript 
 and many useful comments.
 The CFRP laminate mirrors have been developed in collaboration with 
 Mitsubishi Electric Corporation, Communication Systems Center.
 The construction of the telescope was supported by a Grant-in-Aide 
 in Scientific Research of the Japan Ministry of Education, Science, 
 Sports and Cupture (No.~07247103), 
 by the Australian Research Council, 
 and also by the Japanese govermental fund through 
 Institute for Cosmic Ray Research, University of Tokyo.
 AK, SL and TY were supported for this work by JSPS postdoctoral fellowships.

\newpage

\newpage

\newpage

\newpage

\newpage

\newpage

\newpage

\newpage

\newpage

\newpage

\newpage

\end{document}